\documentclass[nofootinbib,aps,11pt]{revtex4-1}
\usepackage{CJK,upgreek,fancyhdr}
\usepackage{graphicx}
\usepackage{hyperref}
\usepackage{cancel}
\usepackage{amssymb}
\usepackage{textcomp}
\usepackage{amsmath}
\usepackage{bm}
\usepackage{times}
\usepackage{epsfig}
\usepackage{color}
\newcommand{\hc}{{\rm h.c.}}
\newcommand{\tr}{{\rm Tr}}
\newcommand{\eV}{{\rm eV}}

\newcommand{\GeV}{{\rm GeV}}
\newcommand{\TeV}{{\rm TeV}}

\newcommand{\br}{{\rm Br}}

\newcommand{\calO}{{\cal O}}

\begin{document}
\begin{CJK*}{GB}{gbsn}
	
\title{\LARGE Confronting the DAMPE Excess with the Scotogenic Type-II Seesaw Model}
\bigskip
\author{Ran Ding $^{1}$}
\email{dingran@mail.nankai.edu.cn}
\author{Zhi-Long Han $^2$}
\email{sps\_hanzl@ujn.edu.cn}
\author{Lei Feng $^3$}
\email{fenglei@pmo.ac.cn}
\author{Bin Zhu $^4$}
\email{zhubin@mail.nankai.edu.cn}
\affiliation{
$^1$ Center for High Energy Physics, Peking University, Beijing 100871, China
\\
$^2$ School of Physics and Technology, University of Jinan, Jinan, Shandong 250022, China
\\
$^3$ Key Laboratory of Dark Matter and Space Astronomy,\\
Purple Mountain Observatory, Chinese Academy of Sciences, Nanjing 210008, China
\\
$^4$ Department of Physics, Yantai University, Yantai 264005, P. R. China}
\date{\today}

\begin{abstract}
 The DArk Matter Particle Explorer (DAMPE) has observed a tentative peak at $E\sim1.4~\TeV$ in the cosmic-ray electron spectrum. In this paper, we interpret this excess in the scotogenic type-II seesaw model. This model extends the canonical type-II seesaw model with dark matter (DM) candidates and a loop-induced vacuum expectation value of the triplet scalars, $v_\Delta$, resulting in small neutrino masses naturally even for TeV scale triplet scalars. Assuming a nearby DM subhalo, the DAMPE excess can be explained by DM annihilating into a pair of triplet scalars which subsequently decay to charged lepton final states. Spectrum fitting of the DAMPE excess indicates it potentially favors the inverted neutrino mass hierarchy. We also discuss how to evade associated neutrino flux in our model.
\end{abstract}

\maketitle

\section{Introduction}\label{IT}

Very recently, the DArk Matter Particle Explorer (DAMPE)  released its  high energy resolution measurement of the cosmic-ray electron spectrum up to $E\sim4.6$ TeV~\cite{Ambrosi:2017wek}. The majority of the spectrum agrees with a smoothly broken power-law model with a spectral break at $E\sim0.9$ TeV, which was previously evidenced by the H.E.S.S. Collaboration~\cite{Aharonian:2008aa}. Remarkably, however, a tentative peak excess at $E\sim1.4$ TeV in the $e^+e^-$ spectrum has been observed, and subsequent analyses showed that the local and global significance of this peak in the DAMPE data reaches about $3.6~\sigma$ and $2.3~\sigma$, respectively~\cite{Huang:2017egk,Fowlie:2017fya}. Since the cooling process of high energy cosmic-ray electrons in the Galactic halo effectively smooths out the spectral features, such a sharp peak indicates there may exist a nearby electron source~\cite{Yuan:2017ysv}. Both astrophysical origin (e.g., an isolated young pulsar) and DM interpretations have been discussed in Ref.~\cite{Yuan:2017ysv}. For the DM interpretation  using model-independent fitting with DM directly annihilating into a pair of standard model (SM) particles, they found that the peak structure can be well fitted for a 1.5 TeV DM particle with the standard WIMP thermally averaged annihilation cross section $\sim3\times10^{-26}{\rm cm}^3 {\rm s}^{-1}$ and annihilation into pure electron or into ${e:\mu:\tau=1:1:1}$ final states, if a nearby DM subhalo is assumed within 1 kpc of the solar system~\cite{Yuan:2017ysv}. In addition, this scenario is
compatible with the constraints from dwarf spheroidal galaxies (dSphs), antiproton and CMB observations~\cite{Yuan:2017ysv}. Since then many relevant studies have been carried out for both simplified and specific (`leptophilic' DM \cite{Bai:2014osa}) model frameworks~\cite{Fan:2017sor,DAMPE}.

In this paper, we apply the scotogenic type-II seesaw model~\cite{Kanemura:2012rj} to account for the DAMPE excess. Unlike the canonical type-II seesaw model, the trilinear interaction between the $SU(2)_L$ triplet scalar $\Delta$ ($L=-2$) and the SM doublet $\Phi$, $\Phi^T i\tau_2\Delta^\dagger \Phi$, is forbidden due to lepton number conservation at the Lagrangian level, while the Yukawa interaction $\overline{L^C_L} (i\tau_2) \Delta L_L$ is still allowed. Meanwhile, a $Z_2$ discrete symmetry is imposed with two $Z_2$-odd scalars, $\chi$ and (singlet, $L=0$) and $\eta$ (doublet, $L=-1$). As a consequence, the lightest scaler, $\chi$, could serve as a DM candidate. The lepton number is spontaneously broken after a $Z_2$-even scalar $\sigma$ (singlet, $L=-1$) develops a vacuum expectation value (VEV). The trilinear interaction $\Phi^T i\tau_2\Delta^\dagger \Phi$ is then induced at one-loop level with $\chi$ and $\eta$ running in the loop. In this way, $v_\Delta$ is naturally suppressed and small neutrino masses generated even for TeV scale triplet scalars.

The triplet scalar $\Delta$ dominantly decays into lepton final states when $v_\Delta\lesssim10^{-4}~\GeV$~\cite{Perez:2008ha}. Therefore, the leptophilic property of DM could be realized through the quartic interaction between $\chi$ and $\Delta$ \cite{Gogoladze:2009gi,Dev:2013hka,Chen:2014lla,Chen:2015cqa,Chen:2017tva,Li:2017tmd}. As a result, the annihilation channel responsible for fitting the DAMPE excess is DM annihilating
into a pair of on-shell triplet mediators, which in turn decay to SM leptons. For nearly degenerate DM and triplet mediator, the $\Delta$ pair is produced almost at rest and each decay final state carries energy of $M_\Delta/2\approx M_\chi/2$. This is equivalent to the $e^+e^-$ spectrum produced by the standard $2\to2$ annihilation process with double the numbers of injection leptons. One thus expects the DAMPE excess could be fitted by setting $M_\chi \sim M_\Delta\simeq3$ TeV. In addition, for a degenerate triplet scalar, its singly-charged and neutral components also produce an accompanying neutrino flux with similar energies of charged leptons, thus can be tested by the existing IceCube data. To avoid this dangerous constraint, we further consider a non-degenerate triplet scalar in which the singly-charged and neutral components are heavier than the doubly-charged component and corresponding neutrino final states are highly suppressed due to the off-shell effect.

The rest of the paper is organized as follows. In Section~\ref{MD}, we briefly review the scotogenic type-II seesaw model and analyse the relevant DM phenomenology. In particular, we give a quantitative estimation of annihilation cross section for each annihilation channel with the off-shell effect included in the non-degenerate case, which allows our model not to  suffer from the neutrino flux constraint. Then in Section~\ref{DP}, we perform detailed spectrum fitting for the DAMPE excess in the non-degenerate case and present benchmarks for both inverted hierarchy (IH) and normal hierarchy (NH) scenarios. Finally, our conclusions are drawn in Section~\ref{CL}.

\section{Model and DM phenomenology}\label{MD}

Embedding DM into the framework of the type-II seesaw mechanism has been widely studied in the literature~\cite{Kanemura:2012rj,Ma:2015xla,Fraser:2015mhb,Guo:2016dzl}. In addition to the SM contents, extra scalar fields with triplet $\Delta$ ($L=-2$), singlet $\chi$ ($L=0$), doublet $\eta$ ($L=-1$) and singlet $\sigma$ ($L=-1$) are introduced in the scotogenic type-II seesaw model~\cite{Kanemura:2012rj}. Moreover, a discrete $Z_2$ symmetry is imposed to stabilize DM, under which only $\chi$ and $\eta$ are arranged to be $Z_2$-odd. In the gauge eigenstates, the components of SM doublet $\Phi$, doublet $\eta$ and triplet $\Delta$ are labeled as
\begin{align}
\Phi=\left(
\begin{array}{c}
\phi^+\\
\phi^0
\end{array}\right),\quad
\eta=\left(
\begin{array}{c}
\eta^+\\
\eta^0
\end{array}\right),\quad \Delta =\left(
\begin{array}{cc}
\delta^+/\sqrt{2} & \delta^{++}\\
\delta^0 & -\delta^+/\sqrt{2}
\end{array}\right),
\end{align}
In this notation, the most general lepton number conservation and $\mathbb{Z}_2$ invariant scalar potential is given by
\begin{eqnarray}\label{Eq:SPT}
V & = & -m_{\Phi}^2 \Phi^\dag \Phi + m_{\Delta}^2 \tr(\Delta^\dag \Delta) - m_{\sigma}^2 \sigma^* \sigma + \frac{1}{2}m_{\chi}^2\, \chi^2 + m_{\eta}^2 \eta^\dag \eta  + \lambda_\Phi (\Phi^\dag \Phi)^2 \\ \nonumber
 &&   + \lambda_{\Delta1} \left[\tr(\Delta^\dag \Delta)\right]^2\!\!  +\lambda_{\Delta2} \tr\left[(\Delta^\dag\Delta)^2\right]\! + \lambda_\sigma (\sigma^*\sigma)^2 + \lambda_\chi\, \chi^4 + \lambda_\eta (\eta^\dag \eta)^2\\ \nonumber
 &&+ \lambda_{\Phi\Delta1} (\Phi^\dag \Phi) \tr(\Delta^\dag \Delta) + \lambda_{\Phi\Delta2} \Phi^\dag \Delta \Delta^\dag \Phi + \lambda_{\sigma \Phi} (\sigma^*\sigma)(\Phi^\dag \Phi) + \lambda_{\chi\Phi} (\Phi^\dag \Phi)\chi^2 \\ \nonumber
 &&  + \lambda_{\eta \Phi1} (\eta^\dag \eta)(\Phi^\dag \Phi) + \lambda_{\eta \Phi 2} (\Phi^\dag \eta)(\eta^\dag \Phi)  + \lambda_{\sigma \Delta} (\sigma^*\sigma) \tr(\Delta^\dag \Delta) + \lambda_{\chi\Delta} \tr(\Delta^\dag \Delta)\chi^2 \\ \nonumber
 &&+\lambda_{\eta\Delta1} (\eta^\dag \eta) \tr(\Delta^\dag \Delta) + \lambda_{\eta\Delta2} (\eta^\dag\Delta \Delta^\dag \eta) + \lambda_{\chi \sigma} (\sigma^*\sigma)\chi^2+\lambda_{\eta\sigma} (\eta^\dag \eta)(\sigma^*\sigma) \\ \nonumber
 &&+\lambda_{\eta \chi} (\eta^\dag \eta)\chi^2+\left[~ \mu_\eta \eta^T (i\tau_2) \Delta^\dag \eta + \lambda_0\,\sigma\chi(\eta^\dag \Phi)+\hc\right]\,.
\end{eqnarray}
The SM electroweak symmetry breaking is triggered by assuming $m^2 > 0$, which leads to a VEV $v\approx246~\GeV$ of the SM doublet $\Phi$. Meanwhile, the parameter $m_\sigma^2$ is also assumed to be positive, to spontaneously break the global lepton number symmetry $U(1)_L$, which induces a VEV $v_\sigma$ of singlet $\sigma$ and a massless majoron $J$~\cite{Bonilla:2015uwa,Wang:2016vfj}. Notice that the majoron would be absorbed when imposing gauged $U(1)_{B-L}$ symmetry~\cite{Wang:2015saa,Wang:2017mcy}. For simplicity, we assume negligible mixing angles between scalars. After spontaneous symmetry breaking,
triplet components in the mass eigenstates are labeled as $H^{\pm\pm}$, $H^\pm$, $H^0$ and $A^0$. Their squared masses are given as,
\begin{eqnarray}
 M^2_{H^{\pm\pm}} &=& m_\Delta^2+\frac{\lambda_{\Phi\Delta1}}{2}v^2
 +\frac{\lambda_{\sigma\Delta}}{2}v_\sigma^2\,, \\ \nonumber
 M^2_{H^{\pm}} &=& m_\Delta^2+\frac{\lambda_{\Phi\Delta1}}{2}v^2 + \frac{\lambda_{\Phi\Delta2}}{4}v^2
 +\frac{\lambda_{\sigma\Delta}}{2}v_\sigma^2\,, \\ \nonumber
 M^2_{H^0,A^0} &=& m_\Delta^2+\frac{\lambda_{\Phi\Delta1}}{2}v^2 + \frac{\lambda_{\Phi\Delta2}}{2}v^2
 +\frac{\lambda_{\sigma\Delta}}{2}v_\sigma^2\,.
\end{eqnarray}
From the above equation, the mass splitting among triplet components yields
\begin{equation}
M^2_{H^\pm}-M^2_{H^{\pm\pm}}=M^2_{H^0,A^0}-M^2_{H^\pm}=\frac{\lambda_{\Phi\Delta2}}{4}v^2,
\end{equation}
which is totally determined by coupling $\lambda_{\Phi\Delta2}$ and triplet components having  degenerate squared mass
\begin{align}
M_\Delta^2=m_\Delta^2+\lambda_{\Phi\Delta1}v^2/2+\lambda_{\sigma\Delta}v_\sigma^2/2,
\end{align}
for vanishing $\lambda_{\Phi\Delta2}$~\cite{Arhrib:2011uy,Han:2015hba,Han:2015sca}. Meanwhile, for non-zero $\lambda_{\Phi\Delta2}$, a mass splitting $|\Delta M|= |M_{H^0,A^0}-M_{H^\pm}|\simeq|M_{H^\pm}-M_{H^{\pm\pm}}|$ could reach $\calO(10)$ GeV under the constraints of electroweak precision tests~\cite{Melfo:2011nx}. In order to respect the limits from LHC direct searches, one requires $M_{\Delta}\gtrsim900~\GeV$~\cite{Aaboud:2017qph}.
The squared masses of $Z_2$-odd scalars are give by
\begin{eqnarray}
M_\eta^2&\simeq& m_\eta^2+(\lambda_{\eta\Phi1}+\lambda_{\eta\Phi2})v^2/2+\lambda_{\eta\sigma}v_\sigma^2/2\,, \nonumber\\
M_\chi^2&\simeq& m_\chi^2 + \lambda_{\chi\Phi} v^2+ \lambda_{\chi\sigma} v_\sigma^2\,.
\end{eqnarray}
Here by choosing $M_\chi<M_\eta$, $\chi$ is the DM candidate.

In this model, the trilinear interaction $\Phi^T i\tau_2\Delta^\dagger \Phi$ is induced at one-loop level and the corresponding effective $\mu$ term is calculated as
\begin{equation}
\mu = \frac{\lambda_0^2 \mu_\eta v_\sigma^2}{64\pi^2(M_\chi^2-M_\eta^2)}\left[1-\frac{M_\chi^2}{M_\chi^2-M_\eta^2}
\ln\frac{M_\chi^2}{M_\eta^2}\right],
\end{equation}
which leads to a VEV $v_\Delta=\mu v^2/M_\Delta^2$ for the triplet. Current precise experimental measurements of $\rho$ parameters limit $v_\Delta$ to less than a few GeV~\cite{Patrignani:2016xqp}. On the other hand, existing constraints from lepton flavor violation processes require $v_\Delta M_\Delta \gtrsim 150~\eV\cdot\GeV$~\cite{Akeroyd:2009nu}. In the following, we take $v_\Delta=1~\eV$ as an illustration. The Yukawa interaction related to neutrino mass generation is given as
\begin{equation}
\mathcal{L}_Y = - Y \overline{L^C_L} (i\tau_2) \Delta L_L + \hc\,,
\end{equation}
where the superscript $C$ denotes charge conjugation and $\tau_2$ is the second Pauli matrix.
The Yukawa matrix $Y$ is complex and symmetric in general, with resulting Majorana neutrino mass matrix
\begin{equation}
\label{Mnu}
M_{\nu} = \sqrt{2} Y v_{\Delta}=V^*m_{\nu}V^{\dag}\,,
\end{equation}
where $m_\nu=\text{diag}(m_1,m_2,m_3)$. $V$ is the PMNS matrix with the following parameterization:
\begin{align}
\!\!\!V=\left(
\begin{array}{ccc}
c_{12}c_{13}&c_{13}s_{12}&e^{-i\delta}s_{13}\\
-c_{12}s_{13}s_{23}e^{i\delta}-c_{23}s_{12}&c_{12}c_{23}-e^{i\delta}s_{12}s_{13}s_{23}&c_{13}s_{23}\\
s_{12}s_{23}-e^{i\delta}c_{12}c_{23}s_{13}&-c_{23}s_{12}s_{13}e^{i\delta}-c_{12}s_{23}&c_{13}c_{23}
\end{array}
\right)\times
\mbox{diag}\left(e^{i\Phi_1/2},1,e^{i\Phi_2/2} \right)\,,
\end{align}
where $s_{ij}=\sin\theta_{ij},~c_{ij}=\cos\theta_{ij}$. $\delta$ and $\Phi_i$ are respectively the Dirac and Majorana CP phases. Using Eq.~(\ref{Mnu}), the Yukawa coupling can be determined as $Y=V^*m_{\nu}V^{\dag}/(\sqrt{2}v_\Delta)$. For illustration, the two Majorana phases are assumed to be zero, and neutrino oscillation parameters are taken to be the best fit values \cite{Esteban:2016qun}:
\begin{eqnarray}\label{nupara}\nonumber
\Delta m_{21}^2 = 7.50\times10^{-5}~\eV^2\,,&\quad&|\Delta m_{31}^2|= 2.524(2.514)\times10^{-3}~\eV^2\,,\nonumber\\
\sin^2\theta_{12}= 0.306\,,&\quad&\sin^2\theta_{23}=0.441(0.587)\,,\nonumber\\
\delta=261^\circ\;(277^\circ)\,,&\quad&\sin^2\theta_{13}=0.02166\;(0.02179)\,,
\label{eq:best}
\end{eqnarray}
where the values in parentheses correspond to the IH case. For  the oscillation parameters given above and $v_{\Delta}=1$ eV, the decay modes and branching ratios of triplet scalar components $H^{\pm\pm},~H^{\pm}$ and $H^0/A^0$ are totally fixed, and are listed in Table~\ref{tab:decay} for both IH and NH scenarios. The decay final states of $H^{\pm\pm}$ yield a fraction of $e^\pm:\mu^\pm:\tau^\pm\simeq 1:0.4:0.7$ for the IH case and $\simeq1:14:18$ for the NH case. This indicates that the IH scenario is preferred by the DAMPE excess, since the spectrum fitting favors electron-rich final states, which is also confirmed by our fitting results in the next section.

\begin{table*}[hbtp]
\begin{tabular}{|c|c|c|c|c|c|c|c|c|c|}
\hline
$H^{\pm\pm}$ & $\br(e^\pm e^\pm)$ & $\br(\mu^\pm \mu^\pm)$ & $\br(\tau^\pm \tau^\pm)$ & \br($e^\pm \mu^\pm)$ & $\br(e^\pm \tau^\pm)$ & $\br(\mu^\pm \tau^\pm)$ & & & \\\hline
IH & $39.44\%$ & $7.09\%$ & $14.36\%$ & $2.63\%$ & $16.18\%$ & $20.30\%$ & & & \\\hline
NH & $0.20\%$ & $24.57\%$ & $34.97\%$ & $2.03\%$ & $3.58\%$ & $34.63\%$ & & & \\\hline
$H^0/A^0$ & $\br(\nu_e \nu_e)$ & $\br(\nu_\mu \nu_\mu)$ & $\br(\nu_\tau \nu_\tau)$ & $\br(\nu_e \nu_\mu)$ & $\br(\nu_e \nu_\tau)$ & $\br(\nu_\mu \nu_\tau)$ & & & \\\hline
IH & $39.44\%\%$ & $7.09\%$ & $14.36\%$ & $2.63\%$ & $16.18\%$ & $20.30\%$ & & &\\\hline
NH & $0.20\%$ & $24.57\%$ & $34.97\%$ & $2.03\%$ & $3.58\%$ & $34.63\%$ & & &\\\hline
$H^\pm$ & $\br(e^\pm \nu_e)$ & $\br(e^\pm \nu_\mu)$ & $\br(e^\pm \nu_\tau)$ & $\br(\mu^\pm \nu_e)$ & $\br(\mu^\pm \nu_\mu)$ & $\br(\tau^\pm \nu_\tau)$ & $\br(\tau^\pm \nu_e)$ & $\br(\tau^\pm \nu_\mu)$ & $\br(\tau^\pm \nu_\tau)$\\\hline
IH & $39.44\%$ & $1.31\%$ & $8.09\%$ & $1.31\%$ & $7.09\%$ & $10.15\%$ & $8.09\%$ & $10.15\%$ & $14.36\%$\\\hline
NH & $0.20\%$ & $1.02\%$ & $1.79\%$ & $1.02\%$ & $24.57\%$ & $17.32\%$ & $1.79\%$ & $17.32\%$ & $34.97\%$\\\hline
\end{tabular}
\caption{The decay modes and branching ratios of the triplet scalar components $H^{\pm\pm}$, $H^0/A^0$ and $H^\pm$ for IH and NH scenarios, with the best fit neutrino oscillation parameters in Eq.~(\ref{eq:best}).}
\label{tab:decay}
\end{table*}

With the above preparation, we now give a simple analysis of DM phenomenology. We implement the complete scotogenic type-II seesaw model in the {\tt FeynRules} \cite{Alloul:2013bka} package with the best-fit oscillation parameters in Eq.~(\ref{nupara}), and apply the {\tt micrOMEGAs4.3.5} package \cite{Belanger:2014vza} to evaluate the DM relic abundance and DM-nucleon scattering cross section. For relic abundance, we adopt the Planck result: $\Omega_{\rm DM} h^2= 0.1199\pm0.0027$ \cite{Ade:2015xua} at the $2\sigma$ confidence level (C.L.). For direct detection constraints, we use the latest spin-independent limits set by the LUX \cite{Akerib:2016vxi}, Xenon1T \cite{Aprile:2017iyp} and PandaX-II \cite{Cui:2017nnn} Collaborations.

Firstly, the DAMPE excess implies that $\chi$ must be leptophilic, which can be  through the annihilation processes:
\begin{equation}
\chi\chi\to H^{++}H^{--},~H^{+}H^{-},~H^0H^0,A^0A^0\,,
\end{equation}
with $H^{\pm\pm}\to\ell^\pm\ell^\pm$ , $H^{\pm}\to\ell^\pm\nu$ and $H^0/A^0\to\nu\nu$ according to the branching ratios in Table~\ref{tab:decay}. In order to guarantee the above annihilation channels are always dominant, one needs the quartic coupling $\lambda_{\chi\Delta}$ to be considerably larger than $\lambda_{\chi\Phi}$. Moreover, the DM-quark interaction is mainly induced by mixing between $\chi$ and SM Higgs, which also demands $\lambda_{\chi\Phi}\lesssim 10^{-2}$ to evade current direct detection constraints~\cite{Li:2017tmd}. As a consequence, leptophilic DM is naturally realized by assuming $\lambda_{\chi\Delta}\gg\lambda_{\chi\Phi}$ in our model.

\begin{figure}
\begin{center}
\includegraphics[width=0.48\linewidth]{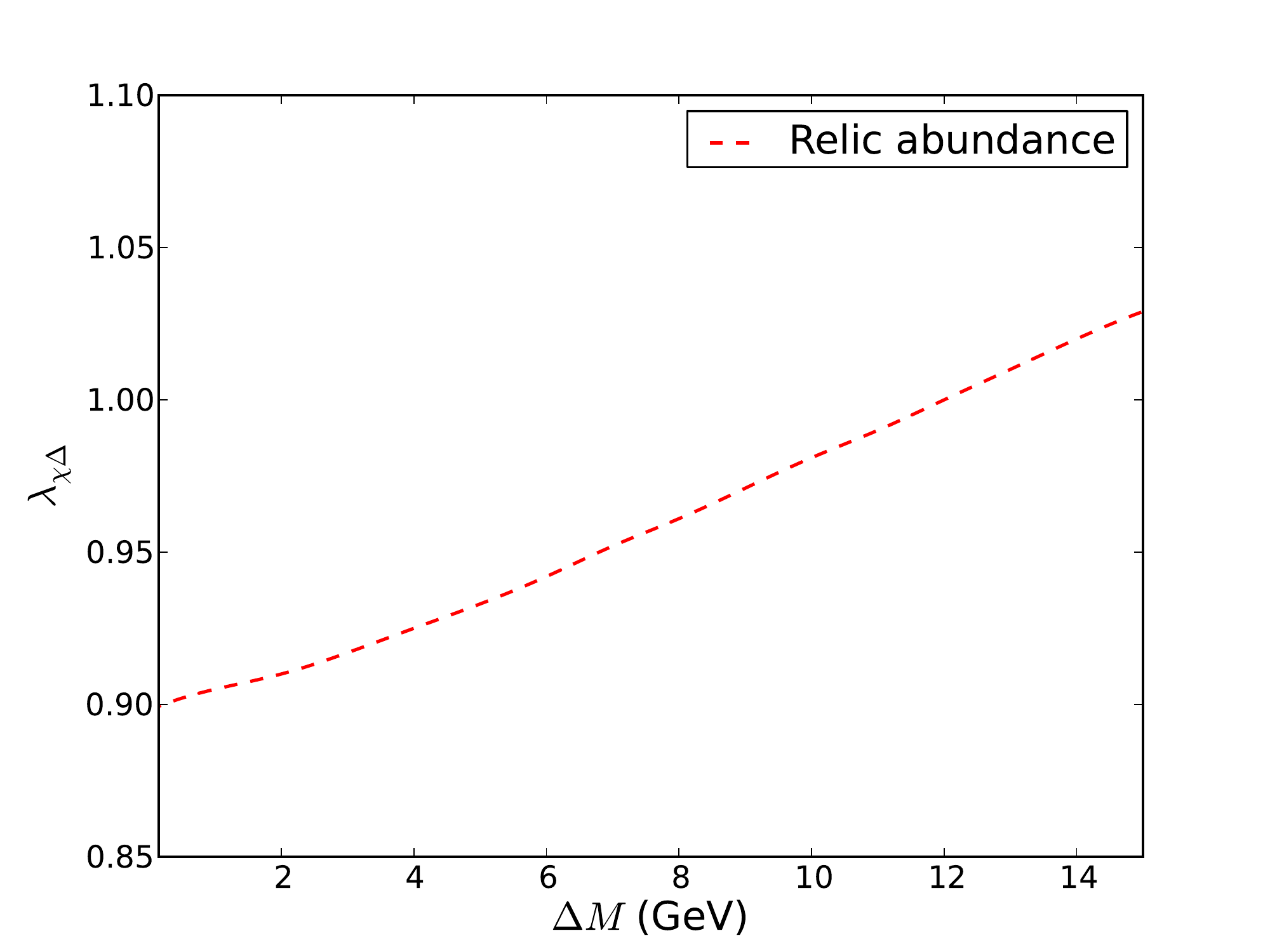}
\includegraphics[width=0.48\linewidth]{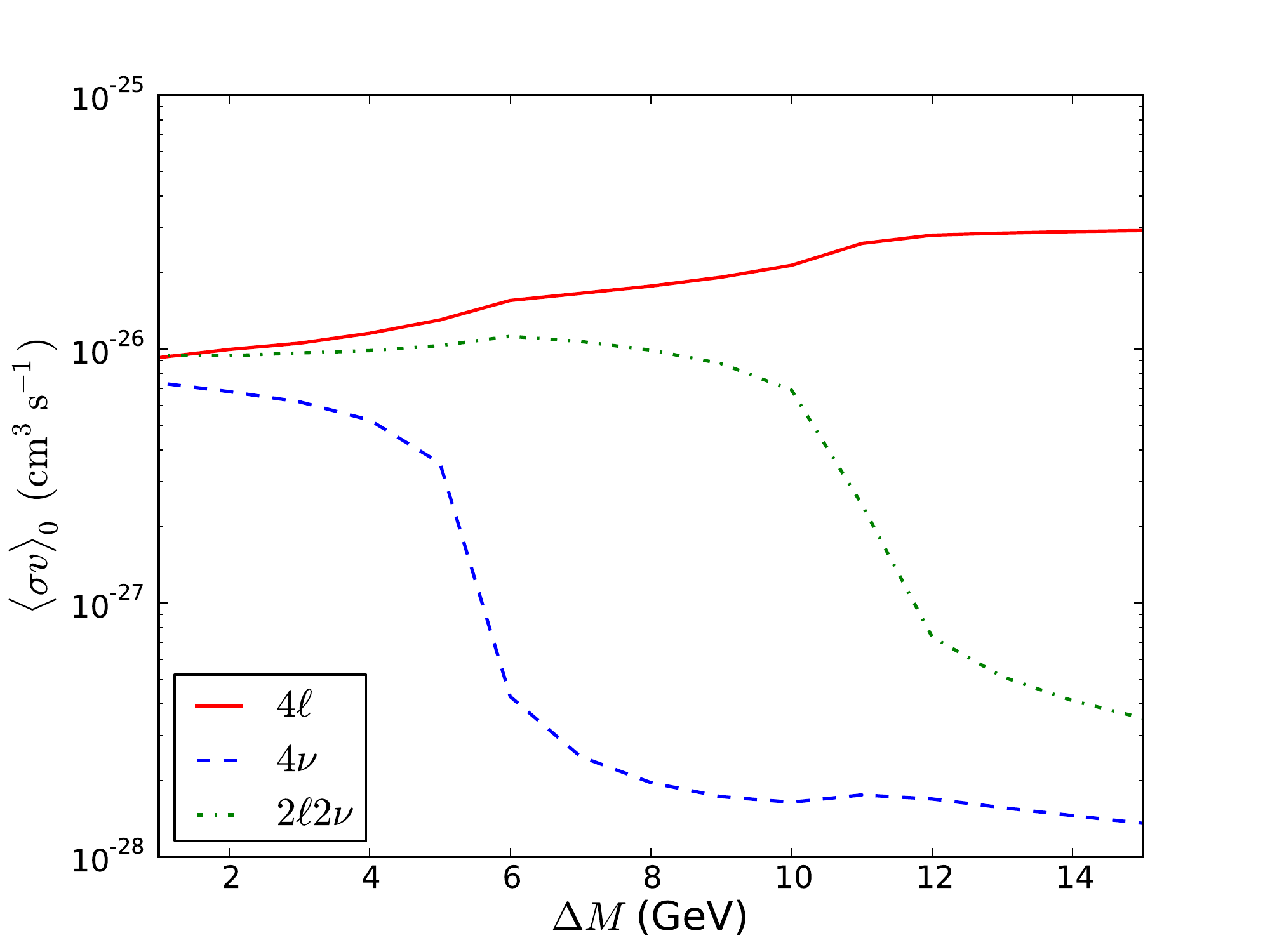}
\end{center}
\caption{Left: The quartic coupling $\lambda_{\chi\Delta}$ required to obtain correct relic abundance as a function of triplet scalar mass splitting $\Delta M$. Right: thermally averaged DM annihilation cross section at present, $\langle\sigma v\rangle_0$, versus $\Delta M$ for different annihilation channels: $\chi\chi \to H^{++}H^{--} \to 4\ell$ (solid red), $\chi\chi \to H^0H^0,A^0A^0 \to 4\nu$ (dashed blue) and $\chi\chi \to H^{+}H^{-}\to 2\ell2\nu$ (dot-dashed green). For both figures, we have fixed the DM masses $M_\chi=3$ TeV and $M_\Delta=2.99$ TeV, respectively.}
\label{fig:cs}
\end{figure}

As we mentioned in the introduction, if the $e^+e^-$ flux produced from DM annihilation originates from left-handed lepton final states, accompanying neutrinos are also produced with comparable flux and similar energies. Such associated neutrinos should in principle carry directional information about the postulated nearby DM subhalo and could be tested at IceCube. According to Ref.~\cite{Zhao:2017nrt}, the 8-year IceCube data is sufficient to identify them with high significance. On the other hand, if the upcoming IceCube observation does not detect such a monochromatic neutrino flux, the explanation based on leptophilic DM and a nearby DM clump may be excluded, which is really a challenge for the usual leptophilic DM model with standard $2\to2$ annihilation processes. However, the situation is different for our model. For the degenerate case in our model, $\chi$ annihilates into $H^{\pm\pm}$, $H^{\pm}$ and $H^0,A^0$ with equal fractions. As a consequence, $50\%$  of the total annihilation cross section contributes to neutrino flux, which is similar to the common leptophilic model, and the IceCube constraint is dangerous. For the non-degenerate case, however, by setting the mass hierarchy $M_{H^0/A^0}=M_{H^{\pm}}+\Delta M=M_{H^{\pm\pm}}+2\Delta M$, the annihilation cross sections related to neutrino final states are highly suppressed due to the off-shell effect. In order to give an accurate estimation, we implement our model in {\tt MadGraph5}~\cite{Alwall:2014hca}, and simulate $\chi$ annihilation at $s=4M^2_\chi+M^2_\chi v^2$ with $v\sim \calO{(10^{-3})}$, which reproduces the conditions of DM annihilation at present. Since we are interested in the cross section, the results from unweighted events are sufficient. In the calculation, we have fixed the DM masses $M_\chi=3$ TeV and $M_{H^{\pm\pm}}=M_\Delta=2.99$ TeV, respectively. Our results are displayed in Fig.~\ref{fig:cs}. Here the right-hand panel presents $\langle\sigma v\rangle_0$ as a function of $\Delta M$ for three annihilation channels: $\chi\chi \to H^{++}H^{--} \to 4\ell$, $\chi\chi \to H^0H^0,A^0A^0 \to 4\nu$ and $\chi\chi \to H^{+}H^{-}\to 2\ell2\nu$, in which the quartic coupling $\lambda_{\chi\Delta}$ for each $\Delta M$ point is evaluated such that corresponding relic abundance is correct, as shown in the left-hand panel. One can see that the behaviors of the three annihilation cross sections are basically determined by the competition between increasing $\lambda_{\chi\Delta}$ and phase space suppression. As a consequence, the $4\ell$ channel monotonically increases with $\Delta M$ just as $\lambda_{\chi\Delta}^2$, since the $H^{\pm\pm}$ pair is always produced on-shell. The $4\nu$ channel monotonically decreases with $\Delta M$ and is highly suppressed when $\Delta M > 5$ GeV due to the off-shell effect. The $2\ell2\nu$ channel increases first then decreases with $\Delta M$, and is eventually highly suppressed when $\Delta M > 10$ GeV. Therefore, for $\Delta M=15$ GeV, annihilation cross sections for both $4\nu$ and $2\ell2\nu$ channels can  safely be neglected.

In addition, one should further consider the secondary neutrino flux resulting from the decays of lepton final states. For the $\mu^+\mu^-$ channel, the IceCube observation from the Galactic center region sets an upper bound with $\langle\sigma v\rangle<9.6\times10^{-23}{\rm cm}^3/{\rm s}$~\cite{Aartsen:2017ulx}. This is much larger than the annihilation cross section required to explain the DAMPE data. In the presence of a subhalo, the corresponding limit is expected to improve by a factor of 2 since the annihilation rate of the subhalo is around two times higher than that of the Galactic center~\cite{Fan:2017sor}. However, this is still far beyond the preferred annihilation cross section in our model. In summary, our scenario is entirely consistent with the current IceCube sensitivity.

Based on the above results, we choose the point of $\Delta M=15$ GeV in Fig.~\ref{fig:cs} as our benchmark (see Table~\ref{tab:bench}) for fitting the DAMPE excess in the next section. Notice that the IH and NH scenarios share the same benchmark since they are only different at decay branching ratios, which does not affect the DM annihilation cross section.

\begin{table*}
\begin{tabular}{|c|c|c|c|c|c|c|c|c|}
\hline
$M_\chi$ (GeV)& $\lambda_{\chi\Delta}$ & $M_\Delta$ (GeV) & $\Delta M$ (GeV)& $\Omega_\chi h^2$ & $\langle\sigma v\rangle_0$ (${\rm cm}^3/{\rm s}$)& $H^{++}H^{--}$  & $H^{0}H^0,A^{0}A^0$ & $H^{+}H^{-}$  \\\hline
$3000$ & $1.03$ & $2990$  & $15$  & $0.119$ & $2.98\times10^{-26}$  & $98.37\%$ & $0.45\%$  & $1.18\%$ \\\hline
\end{tabular}
\caption{The DM information for the IH/NH benchmark in the non-degenerate case. Here $\langle\sigma v\rangle_0$ denotes the thermally averaged DM annihilation cross section at present. The last three columns list the relative contributions of annihilation channels $\chi\chi \to H^{++}H^{--} \to 4\ell$, $\chi\chi \to H^{0}H^0,A^{0}A^0 \to 4\nu$ and $\chi\chi \to H^{+}H^{-}\to 2\ell2\nu$, respectively.}
\label{tab:bench}
\end{table*}

\section{Spectrum Fitting of DAMPE Excess}\label{DP}

In this section, we give a spectrum fitting for the DAMPE excess based on the benchmark in Table~\ref{tab:bench}, by taking into account the contribution of a nearby DM subhalo. Before showing our results, we briefly describe the fitting prescription. The differential number density $f(t,\vec{x},E)=dN/dE$ obeys the diffusion-loss equation~\cite{Cirelli:2010xx}:
\begin{align}
\frac{\partial }{\partial t}f(t,\vec{x},E)-\vec{\nabla}\cdot\left[K(\vec{x},E)\vec{\nabla} f(t,\vec{x},E)\right]-\frac{\partial}{\partial E}\left[ b(\vec{x},E)f(t,\vec{x},E)\right]=Q(\vec{x},E)\,.
\label{eq:diffusion0}
\end{align}
For the steady-state case, the diffusion equation is reduced to,
\begin{align}
-\vec{\nabla}\cdot\left[K(E)\vec{\nabla}f(\vec{x},E)\right]-\frac{\partial}{\partial E}\left[ b(E)f(\vec{x},E)\right]=Q(\vec{x},E)\,,
\label{eq:diffusion}
\end{align}
which only keeps the space diffusion and energy loss effects. Here the function $Q(\vec{x},E)$ is the source term. $K(E)=K_0(E/E_0)^\delta$ is the diffusion coefficient with slope $\delta$, which depends on the Galactic diffusion cylinder with height $2L$. Here we choose the propagation parameters as $K_0=0.1093$ ${\rm kpc}^2~{\rm Myr}^{-1}$, $\delta=1/3$ and $L=4~{\rm kpc}$. $b(E)=E^2/(E_0\tau_E)$ is the positron loss rate due to synchrotron radiation in the Galactic magnetic field and inverse Compton scattering with CMB photon and stellar light. The typical loss time $\tau_E=10^{16}~{\rm s}$ and $E_0=1$ GeV. Eq.~(\ref{eq:diffusion}) can be solved by using the Green function method, and the general solution is expressed as,
\begin{align}
f(x_{\odot},E)=\int^{m_{\rm DM}}_E dE_s\int d^3\vec{x}_sG(\vec{x}_{\odot},E;\vec{x}_s,E_s)Q(\vec{x}_s,E_s)\,,
\label{eq:Green}
\end{align}
where $f(x_{\odot},E)$ is the $e^\pm$ number density at the earth with energy $E$, and $E_s$ denotes $e^\pm$ energy at the source position $\vec{x}_s$. From the above equation, the differential flux of $e^{\pm}$ is evaluated as
\begin{align}
\Phi_{e^\pm}(x_{\odot},E)=\frac{v_{e^\pm}}{4\pi}f(x_{\odot},E)\,.
\end{align}
In the above equation, the velocity of $e^\pm$ approximately yields $v_{e^\pm} = c$. For a nearby DM subhalo at position $x_{\rm sub}$, the source term is given as
\begin{align}
Q(\vec{x},E) = \frac{\langle \sigma v \rangle_0}{2M_\chi^2}\int \rho^2(r)dV \delta^3(\vec{x} -\vec{x}_{\rm sub})\,,
\label{eq:source}
\end{align}
where $\langle \sigma v \rangle_0\simeq\langle \sigma v \rangle_0(H^{++}H^{--})$ for our benchmark, and $\rho(r)$ is the density profile of the subhalo. We adopt the NFW density profile~\cite{Navarro:1995iw,Navarro:1996gj}:
\begin{align}
\rho(r)=\frac{\rho_s}{(r/r_s)(1+r/r_s)^2}\,,
\end{align}
for both Galactic halo and subhalo. For the Galactic halo, the
two parameters $\rho_s$ and $r_s$ are normalized by the local density $\rho_{\odot}$ and distance from Galactic center to  Sun $R_{\odot}$, which are respectively taken as $\rho_{\odot}=0.4$ ${\rm GeV}{\rm cm}^{-3}$ and $R_{\odot}=8.5$ kpc. For the nearby subhalo, $\rho_s$ and $r_s$ are determined by its mass $M_{\rm sub}$ (after tidal stripping)~\footnote{For the calculation of $M_{\rm sub}$, see the Appendix of Ref.~\cite{Yuan:2017ysv}.}. With its distance to the solar system $d$, the features of the nearby subhalo are characterized by free parameters $(d,~M_{\rm sub})$. We use {\tt micrOMEGAs} to evaluate the $e^+e^-$ spectrum
produced by DM annihilation in the Galactic halo, and numerically solve the integral equation Eq.~(\ref{eq:Green}) with the source term in Eq.~(\ref{eq:source}) to calculate the subhalo contribution. In addition, the background flux coming from various astrophysical sources also need to taken into account. We use the {\tt GALPROP} package~\cite{Moskalenko:1997gh,Strong:1998pw} here and perform $\chi^2$ analysis to obtain the best-fit astrophysical background, which yields $\chi^2_{\rm bkg}=108.04$.

\begin{figure}
\begin{center}
\includegraphics[width=0.8\linewidth]{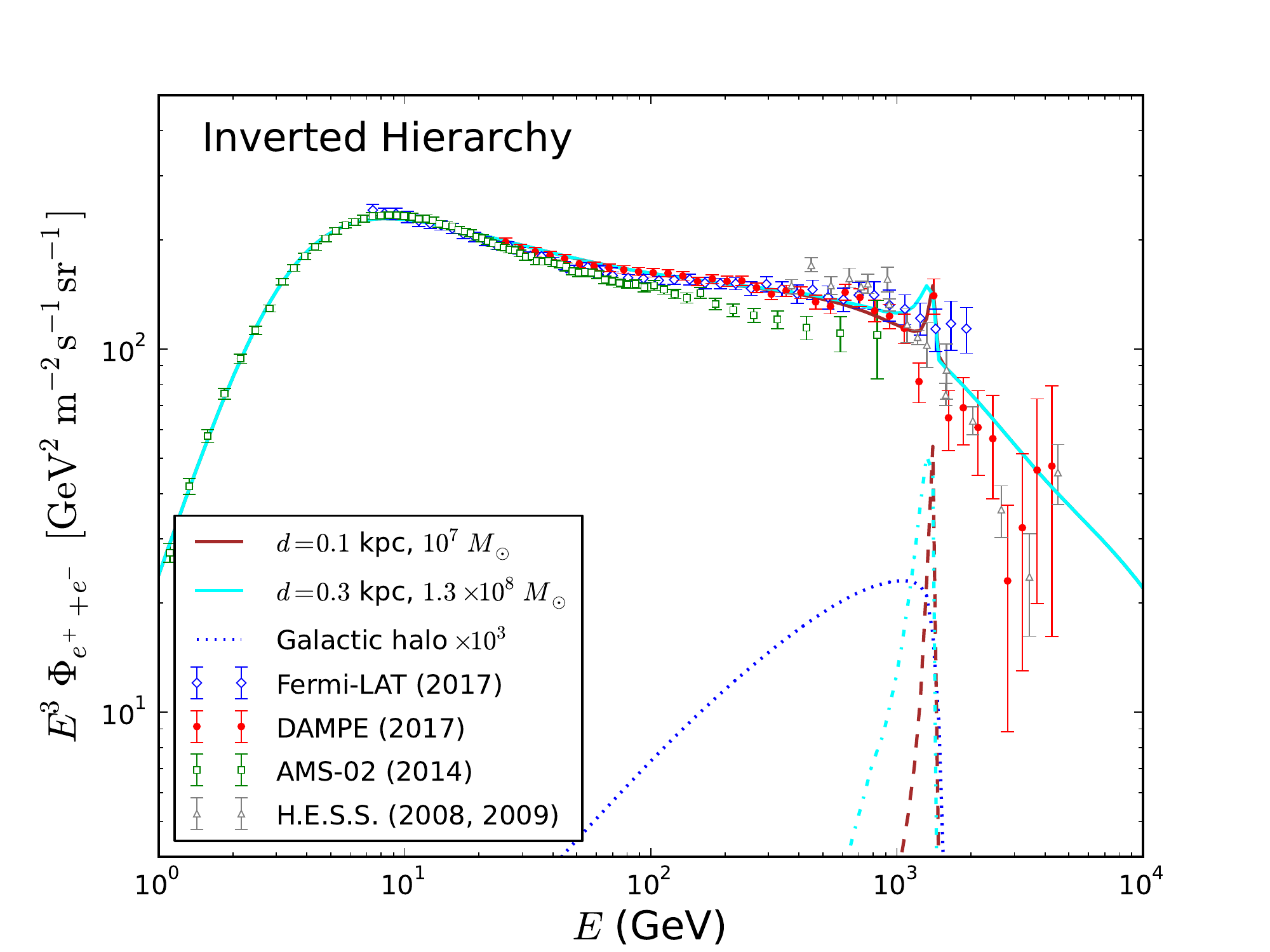}
\end{center}
\caption{The $e^+e^-$ spectrum for the IH benchmark. The DAMPE data is shown by the red points. The dotted blue  line denotes the Galactic halo contribution (multiplied by $10^3$). The dashed green  and dot-dashed  cyan lines respectively denote the contributions of a nearby subhalo with parameters $(d,~M_{\rm sub})=(0.1~{\rm kpc},~10^7~M_{\odot})$ and $(0.3~{\rm kpc},~1.3\times10^8~M_{\odot})$. The corresponding total fluxes (background + Galactic halo + subhalo) are shown by solid lines with the same colors. Here we take the solar modulation potential as $\Phi_{\odot}=700$ MV for illustration. For comparison, we also present measurements from the AMS-02~\cite{Aguilar:2014mma}, Fermi-LAT~\cite{Abdollahi:2017nat} and H.E.S.S.~\cite{Aharonian:2008aa} experiments. The error bars of DAMPE, AMS-02 and Fermi-LAT include both systematic and statistical uncertainties.}
\label{fig:benchIH}
\end{figure}

\begin{figure}
\begin{center}
\includegraphics[width=0.8\linewidth]{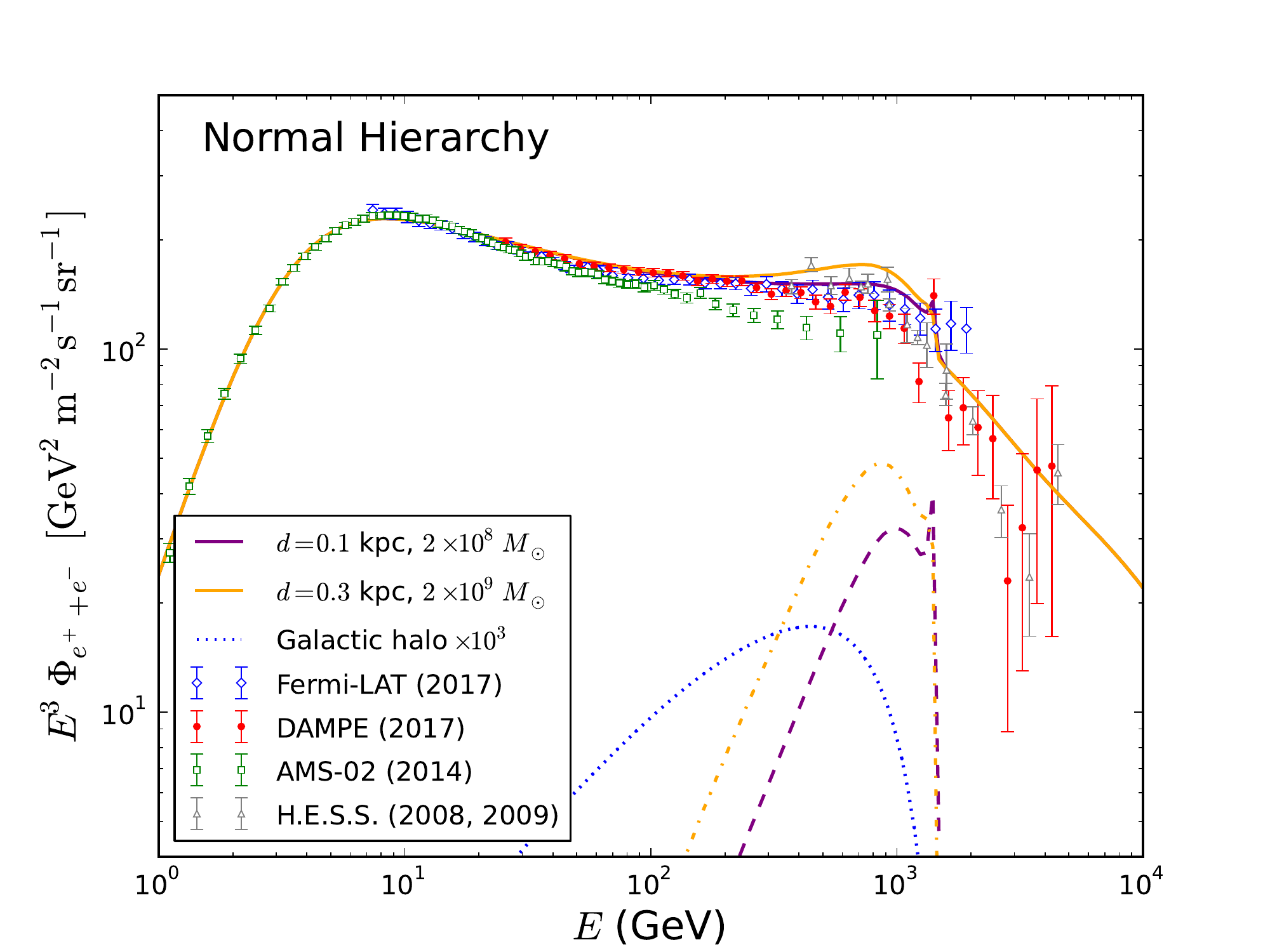}
\end{center}
\caption{The $e^+e^-$ spectrum for the NH benchmark. The DAMPE data is shown by the red points. The dotted blue  line denotes the Galactic halo contribution (multiplied by $10^3$). The dashed purple  and  dot-dashed orange lines respectively denote the contributions of a nearby subhalo with parameters $(d,~M_{\rm sub})=(0.1~{\rm kpc},~2\times10^8~M_{\odot})$ and $(0.3~{\rm kpc},~2\times10^9~M_{\odot})$. The corresponding total fluxes (background + Galactic halo + subhalo) are shown by solid lines with the same colors. Here we take the solar modulation potential as $\Phi_{\odot}=700$ MV for illustration. For comparison, we also present measurements from the AMS-02~\cite{Aguilar:2014mma}, Fermi-LAT~\cite{Abdollahi:2017nat} and H.E.S.S.~\cite{Aharonian:2008aa} experiments. The error bars of DAMPE, AMS-02 and Fermi-LAT include both systematic and statistical uncertainties.}
\label{fig:benchNH}
\end{figure}

The resulting $e^+e^-$ spectrum for our benchmark is presented in Fig.~\ref{fig:benchIH} for the IH scenario, and in Fig.~\ref{fig:benchNH} for the NH scenario. From these two figures, one can see that the morphology of the DAMPE excess potentially favors the IH scenario, which could well fit the tentative peak for subhalo parameters $(d,~M_{\rm sub})=(0.1~{\rm kpc},~10^7~M_{\odot})$ or $(0.3~{\rm kpc},~1.3\times10^8~M_{\odot})$. This is because for the IH scenario, $\chi$ predominately annihilates into electron final states $(e:\mu:\tau\simeq 1:0.4:0.7)$, which have a sharp prompt spectrum. As a consequence, the diffusion and energy loss effects are not significant for a nearby source. For the NH scenario, however, $\chi$ predominately annihilates into tau final states $(e:\mu:\tau\simeq1:14:18)$, which have a much broader prompt spectrum compared with electrons. The spectrum is further smeared via diffusion and energy loss, thus resulting in a worse fitting. For instance, for our benchmark with subhalo parameters $(d,~M_{\rm sub})=(0.1~{\rm kpc},~2\times10^8~M_{\odot})$ or $(0.3~{\rm kpc},~2\times10^9~M_{\odot})$, it cannot match the tentative peak structure in the DAMPE data.

Before ending this section, we mention that accompanying $\gamma$-ray photons are also produced due to the internal bremsstrahlung process and the decay of the charged lepton final states. It is also necessary to check whether such $\gamma$-ray emission can be detected or constrained by current observations. The $\gamma$-ray flux from the nearby DM subhalo contribution has been estimated in Ref.~\cite{Yuan:2017ysv} for the $e^+e^-$ and ${e:\mu:\tau=1:1:1}$ annihilation channels, with typical integral radius within $1^\circ$ and subhalo distance $d=0.1/0.3$ kpc. They found that in all cases, corresponding $\gamma$-ray fluxes are below the 10-year point source sensitivity of Fermi-LAT observations. This estimation is also applicable to our model.

\section{Conclusion}\label{CL}

We have interpreted the recent DAMPE excess in the framework of the scotogenic type-II seesaw model which relates neutrino masses and scalar singlet DM $\chi$ at one-loop level. By assuming a nearby DM subhalo, the DAMPE excess can be fitted for our benchmark in the non-degenerate case and IH scenario, where DM $\chi$ predominately annihilates into a pair of on-shell $H^{\pm\pm}$ mediators, which then decay to electron rich final states. In this case, associated neutrino final states are highly suppressed due to the off-shell production of $H^0,A^0$ and $H^{\pm}$ mediators. This advantage means our model does not suffer from the limits on the neutrino flux from IceCube. In addition, the $\gamma$-ray flux in our model is also below the current Fermi-LAT sensitivity. Finally, the lepton flavor structure in our model which produces the primary $e^+e^-$ flux is tightly related to the neutrino  oscillation data. We find that the spectrum fitting of the DAMPE excess potentially favors the IH for neutrino mixing.

\textbf{Acknowledgements.} We thank Qiang Yuan for help with the DAMPE spectrum fitting, and Yue Zhao for useful discussions on neutrino flux constraints.


\clearpage

\end{CJK*}
\end{document}